\documentstyle[aps,preprint,prl]{revtex}
\begin{document}
\flushbottom

\widetext
\draft
\title{The Solar Neutrino Puzzle: An Oscillation Solution\\
with Maximal Neutrino Mixing }
\author{Anthony J. Baltz$^a$, Alfred Scharff Goldhaber$^b$, and Maurice
Goldhaber$^a$}
\address{
$^a$Physics Department,
Brookhaven National Laboratory,
Upton, New York 11973}

\address{
$^b$Institute for Theoretical Physics,
State University of New York,
Stony Brook, NY 11794-3840}
\date{\today}
\maketitle

\def\thepage{\arabic{page}}
\makeatletter
\global\@specialpagefalse
\ifnum\c@page=1
\def\@oddhead{Draft\hfill To be submitted to Phys. Rev. C}
\else
\def\@oddhead{\hfill}
\fi
\let\@evenhead\@oddhead
\def\@oddfoot{\reset@font\rm\hfill \thepage \hfill}
\let\@evenfoot\@oddfoot
\makeatother

\begin{abstract}
If, as suggested by the SuperKamiokande results, $\nu_{\mu}$ and $\nu_{\tau}$
are maximally
and ``rapidly'' ($\Delta m^2 \approx 2.2 \times 10^{-3} {\rm (eV)^2}$) mixed,
this alone determines the mapping from current to mass eigenstates up to one
rotation angle $\theta$ mixing $\nu_e$, ``more slowly'', with an equal combination
of $\nu_{\mu}$ and $\nu_{\tau}$. For ${\rm
sin}2\theta=1$, the resulting minimal number of free
parameters, yet maximal mixing, shows agreement between extant observations
of solar neutrinos and predictions by the standard solar model with
minor modifications.
\\
{\bf PACS: {14.60.Pq, 13.10.+q, 25.30.Pt}}
\end{abstract}

\makeatletter
\global\@specialpagefalse
\def\@oddhead{\hfill}
\let\@evenhead\@oddh

When Kajita\cite{kaj} reported at Neutrino '98 evidence for oscillation of
atmospheric neutrinos with 
$\Delta m^2 \approx 2.2 \times 10^{-3} {\rm (eV)^2}$ 
and large mixing, probably between
$\mu$ and $\tau$ neutrinos,
the conceptual landscape for discussion of neutrino mixing changed
dramatically.  The simplest interpretation consistent with this result is
that there is maximal mixing between $\nu_\mu$ and $\nu_\tau$ and negligible
mixing with $\nu_{\rm e}$.  This remarkable conclusion leads to an important
application
in that other great arena, where neutrino oscillations have long been
suspected but have so far eluded definitive proof, solar neutrinos.
We do that here by assuming that the one parameter left
free by the new result, the amount of mixing of $\nu_{\rm e}$, also is maximal,
and then comparing deductions from that assumption with current
observations, as
well as predicting consequences for possible future observations.

At the very beginning of particle-physics attacks on the deficit in neutrinos
arriving from the sun, as compared with expectations from the standard
solar model [SSM], see Ref.\cite{bah}, it was clear that maximal mixing of
$\nu_{\rm e}$ and $\nu_\mu$ would go a very long way in solving the puzzle.
However, before the new SuperKamiokande result,there were
strong reasons to be cautious about such a hypothesis: (1){\it
Phenomenology:} The
nearest analogue, the CKM matrix mapping quark electroweak current
eigenstates to
mass eigenstates shows mixing that is small between adjacent generations
and very
small between the highest and lowest generations
\cite{ckm}.  (2)  {\it Theory:} The
widely accepted seesaw mechanism \cite{seesaw} for neutrino masses also
suggests
small mixing angles\cite{blud}.  (3) {\it Superfluity:} The MSW effect
(so called after Mikheyev, Smirnov and Wolfenstein) seemed able to
give a rigorous explanation for the solar neutrino deficit even with small
mixing,
provided the relevant values of $\sin 2 \theta$ and
$\Delta m^2$ for $\nu_{\rm e}$ mixing
lie in a limited range. (4) {\it Esthetics:}  Once one knew that there were
three
generations of neutrinos, why should $\nu_{\rm
e}$ be linked strongly with just one other generation?  This last objection
could be met
by the complete three-generation-maximal mixing as discussed by several
authors\cite{hps}, but this scenario suggests too small a reduction.
Thus
there was neither experimental evidence nor theoretical motivation for large,
much less maximal, mixing.

The ideal assumption of maximal mixing between $\nu_\mu$ and $\nu_\tau$ for
small values of $L / E$ (earth's dimensions and GeV energies) has the immediate
consequence that
by suitable phase convention choices one mass eigenstate
 $|\nu_3 \! >$ may be
written (as ilustrated in Fig. 1)
\begin{equation}
|\nu_3 \! >=(|\nu_\mu \!  > + |\nu_\tau \!   \! >)/\sqrt{2} \ .
\end{equation}
The most general form for the two other mass eigenstates then becomes
\begin{equation}
|\nu_1 \! >={\rm cos} \theta  |\nu_{\rm e} \!  > + {\rm sin}
\theta |\nu'
\! >
\end{equation} and
\begin{equation}
|\nu_2 \! >=-{\rm sin} \theta |\nu_{\rm e} \!  > + {\rm cos} \theta |\nu'
\! > \ ,
\end{equation}
with
\begin{equation}
|\nu' \! >=(|\nu_\mu \!  > - |\nu_\tau \!   \! >)/\sqrt{2}
\end{equation} and
\begin{equation}
|m_3^2 - m_2^2| \approx 2.2 \times 10^{-3} {eV^2} >> |m_2^2-m_1^{2}| \ .
\end{equation}

Thus, the issue of $\nu_e$ mixing becomes a two-state problem, with
the only change from what might have been done years ago being that $\nu'$
takes the place of $\nu_\mu$ as the mixing partner.  (Note that $\nu'$ is
neither a flavor nor a mass eigenstate.)
The combination of
the atmospheric SuperKamiokande result and the maximal mixing hypothesis for
 $\nu_e$ uniquely
specifies the mapping from the current eigenstates to the mass
eigenstates.
Note that because we have been allowed to choose the mapping as completely
real, no CP violation arises in the mixing.  For that, a necessary
requirement would be that each of the three mass eigenstates involves all
of the current eigenstates.

It follows from the hypothesis that oscillations of $\nu_e \longleftrightarrow
\nu_{\mu}$ as well as $\nu_e \longleftrightarrow
\nu_{\tau}$ should be negligble for
atmospheric neutrinos.  This is compatible with present observations
by SuperKamiokande (see\cite{kaj}), but the conclusion depends on the absolute
number of
atmospheric
$\nu_e$'s predicted.  It will be interesting to see whether the results of
calculations
which take account of the different paths of pions and muons in the Earth's
magnetic field will affect this conclusion(see Gaisser \cite{gai98}).

Compared to the expectations from the published Standard Solar Model (SSM)
\cite{bp},
the various detectors for solar neutrinos (Homestake\cite{hom},
GALLEX\cite{gal}, SAGE\cite{sag}, Kamiokande and
SuperKamiokande\cite{suk}) have shown deficiencies, often interpreted as due to
matter-induced resonant oscillations in the sun (the MSW effect), where the
electron
neutrinos change flavor to a state for which the detectors are
insensitive or less sensitive.
These oscillations are characterized by a mixing angle
$\theta$ and the difference of squared masses $\Delta m^2$ = $m_2^2 - m_1^2$,
where $m_1$ and $m_2$ refer to mass eigenstates.  A mixed state
propagates through the vacuum with oscillation length $L_v$\cite{bah}
\begin{equation}
L_v = 2.48 \times 10^{-3} {E_{\nu} (MeV) \over \Delta m^2 (eV)^2} km.
\end{equation}

Various solutions for the parameters
$\theta$ and $\Delta m^2$ are compatible with the data.  The MSW effect
yields possible central solutions
$\Delta m^2 = 5.1\times 10^{-6}{\rm(eV)^2},
\sin^2 2 \theta = 8.2\times 10^{-3}$, and $\Delta m^2 = 1.6\times 
10^{-5}{\rm(eV)}^2,
\sin^2 2 \theta = 0.63$ (see Hata and Langacker\cite{hata}).
Since matter enhanced effects become unimportant as
$\sin 2 \theta \rightarrow 1$, 
the MSW mechanism
is neither needed nor operative for maximal mixing.
The special case of a ``just-so'' vacuum solution has been
discussed by Krastev and Petcov\cite{kp2}.  For a recent review of
the entire current solar neutrino situation see e.g. Berezinsky\cite{ber}.

Let us assume that the neutrino deficiencies
found are
partially due to oscillations of electron neutrinos to different flavors,
and partially due to an
overestimate of the last, and probably weakest, link
in the main neutrino chain of the SSM, viz. the emission intensity of $^8$B
neutrinos.  The minimum required deficiency in
emission is obtained for maximal neutrino mixing.
If a detector integrates over a sufficient range of energies
and/or a sufficient range of distances, phase averaging leads, after many
oscillations, to a reduction of the
expected signal by a factor two.  Since the number of
$^8$B neutrinos is found by SuperKamiokande to be
less than half of the SSM value\cite{suk} the assumed vacuum solution
would imply that there
is a deficit in emission of $^8$B neutrinos, compared with
expectations from the SSM.

For the chemical detectors ($^{37}$Cl and $^{71}$Ga) the maximal mixing
vacuum solution would lead for phase averaging to a halving of the expected
neutrinos
detected as the experiments are not sensitive to muon or tau neutrinos.
In the water \^Cerenkov detectors
muon or tau neutrinos are both detected at a rate reduced to about $14.7\%$ of
the detection rate for electron neutrinos, when averaged over the part of the
spectrum detected by SuperKamiokande.
Assuming the rate of $^8$B neutrinos emitted by the sun to be $(1 - x)$ times
the value predicted by the SSM,
the ratio $R(^8$B) of electron recoils observed by SuperKamiokande,
relative to the
expectation from the SSM without oscillations, can be written as
\begin{equation}
R(^8{\rm B}) = {1 \over 2} \times (1 + 0.147) \times (1 - x) = 0.368\ {\rm or }
\ (0.474)
\end{equation}
giving a reduction $x \sim 0.36$ or $(0.17)$ for the $^8$B
neutrinos, when the 1995\cite{bp} or (1998\cite{bp98}) version of the
Bahcall Pinsonneault SSM is considered.

The reduction of $^8$B from the SSM predictions is shown
in Figure 2.  This allows us, as explained in the legend, to test the
consistency of our model with the results obtained by the $^{37}$Cl and
$^{71}$Ga
experiments.  For BP95 SSM we find a $36\%$ reduction of the $^8$B
neutrinos emmited by the sun.  This leads to a prediction in agreement with
the $^{71}$Ga results but misses the $^{37}$Cl result, overesimating it.
The recently revised
SSM (BP98) makes use of a $^7$Be (p, $\gamma$) $^8$B cross section reduced
by 15\% from BP95 and of revised solar dynamics, that reduce the
$^8$B neutrino flux to 78\% of of that prediced by BP95.  Our maximal mixing
model then calls for only a 17\% reduction of $^8$B neutrino flux from
BP98.  Again our prediction is in
agreement with the $^{71}$Ga results, but misses the $^{37}$Cl result
similarly, by overestimating it.

The solution of maximal mixing, with a reduction in the emission of $^8$B
neutrinos, is consistent
with a large range of possible values of $\Delta m^2$.  The value of
$\Delta m^2$ must be large enough to achieve phase averaging of the
oscillations
for the various neutrino sources in the sun.  At a value of
$5-9 \times 10^{-11}$ (eV)$^2$ there is a ``just so'' vacuum oscillation
solution relying on the oscillation phase\cite{hata},
corresponding to several ($\sim 2-4$) full wave length
oscillations on the way from sun to earth (mean distance $= 1.49 \times 10^8$
km).
The vacuum oscillation formula for survival of an electron neutrino with
maximal mixing is\cite{bah}
\begin{equation}
P(\nu_e) =1 -  \sin^2 2 \theta \sin^2 {\pi L \over L_v},
\end{equation}
where $L$ is the distance from the sun and $L_v$ is given by Eq.(1).
For the
scattering by electrons of the
monoenergetic $^7$Be neutrinos, which BOREXINO intends to observe, the
detection rate (normalized to unity for
no oscillations) becomes
\begin{equation}
R(^7{\rm Be}) = 1 - 0.79 \sin^2 2 \theta
\sin^2 {\pi \Delta m^2 (eV)^2  L (km) \over (0.862)  2.48 \times 10^{-3}},
\end{equation}
where the muon or tau neutrino scattering relative to electron neutrino
scattering at 0.862 MeV is 0.21\cite{bah}.  As Krastev and
Petcov\cite{kp2}, for example,
point
out, there is a large change in the $^7$Be electron neutrino flux over the
year for the ``just-so'' vacuum solutions
due to the change in phase of the order of $\pi / 2$ in a year
brought about by the $\pm 1.67\%$ yearly orbital
variation from the mean distance of the sun to the earth.
GALLEX, where individual experiments represent averages in neutrino absorption
over serveral weeks, did not observe a
seasonal effect\cite{sea}.  For a value of
$\Delta m^2 > \sim 10^{-9}$ (eV)$^2$ the oscillation would go
through many complete phases in a year and one would attain the region where
our phase averaged vacuum mixing model would hold for the $^{71}$Ga
detectors. However Suzuki\cite{suk} reports a hint of a distortion of the
$^8$B
spectrum.  If this preliminary result should be confirmed
we would have to reconsider some of our conclusions.
But for the present, we take our solution
to approximately span the mass region $10^{-9} < \Delta m^2 <<
0.9 \times 10^{-3}$, using the CHOOZ upper limit\cite{chooz}.

The variation in orbital
distance ($5 \times 10^6$ km) may be compared to the average source size of the
shell in the sun whence the $^7$Be neutrinos originate
($\sim 10^5$ km)\cite{bp}.
Since the phase change in a year is $\sim 50$ times the phase averaging due
to the source size, it dominates on a yearly average.  However, if one had
sufficient statistics to measure the $^7$Be intensity on, say, a daily basis,
then the change
in phase from day to day due to the earth's orbit would be of the same order of
magnitude as the phase variation (averaging) at the source, thus allowing
an island of $\Delta m^2$ at $\sim 10^{-8}$ (eV)$^2$ to be explored.

We summarize here some experimental consequences of our
solution which can be tested
by the existing or soon to be completed neutrino detectors:

(1) There is no distortion of the $^8$B neutrino spectrum of the kind demanded
by an MSW effect in the sun.  However, see the remark above relating to
Suzuki's report at Neutrino '98.

(2) The deficit of $\sim 36\% $ ($\sim 17\% $) for $^8$B neutrinos can
be tested when neutral current interactions are studied at SNO.

(3) Our value for $R(^7$Be) can be tested at BOREXINO.

(4) There is no seasonal effect for $^7$Be neutrinos, other than the small
variation due to the $1/$r$^2$
effect, where r is the sun-earth distance.

(5) There should be no day-night effect (see \cite{bw} arising from
matter oscillations in the sun.  A different
day-night effect where solar electron neutrinos interact with the core of the
Earth has recently been proposed by several authors (parametric resonance)
\cite{sm}.

We should like to thank M. Diwan, W. J. Marciano, P. G. Langacker,
S. T. Petcov,
S. P. Rosen, and A. Yu. Smirnov for valuable discussions.

(While this paper was being completed our attention was drawn to a
manuscript posted by V. Barger, S. Pakvasa, T. J. Weiler, and
K. Whisnant\cite{bpww}, which has some similar considerations.)

\begin{figure}
\caption[Figure 1]{The figure shows in perspective the three-dimensional
principal axis
transformation from the current eigenstates to the mass eigenstates.
First, the system is rotated $45^o$ about the $\nu_e$ direction, thus taking
the original $\nu_\tau$ direction into the final $\nu_3$ direction.
Secondly, the system is rotated $45^o$ about the $\nu_3$ direction,
taking the original $\nu_e$ direction into the final $\nu_1$ direction.}
\end{figure}
\begin{figure}
\caption[Figure 2]{Rates observed by the solar neutrino detectors compared
with rates predicted for maximal neutrino mixing
as a function of the reduction of the $^8$B neutrino flux in the sun from the
predictions of the SSM BP95 (heavy dot-dashed line) and BP98 (faint dot-dashed
line) are shown in all three boxes.  Note that the vertical scale is
logarithmic for all three plots.  Heavy horizontal lines represent the
experimental values, with dashed lines the errors.  Errors shown on the
right side for BP95 are similar to those for BP98 (not shown).
The $^{71}$Ga data are
an average of the GALLEX and SAGE data.}
\end{figure}
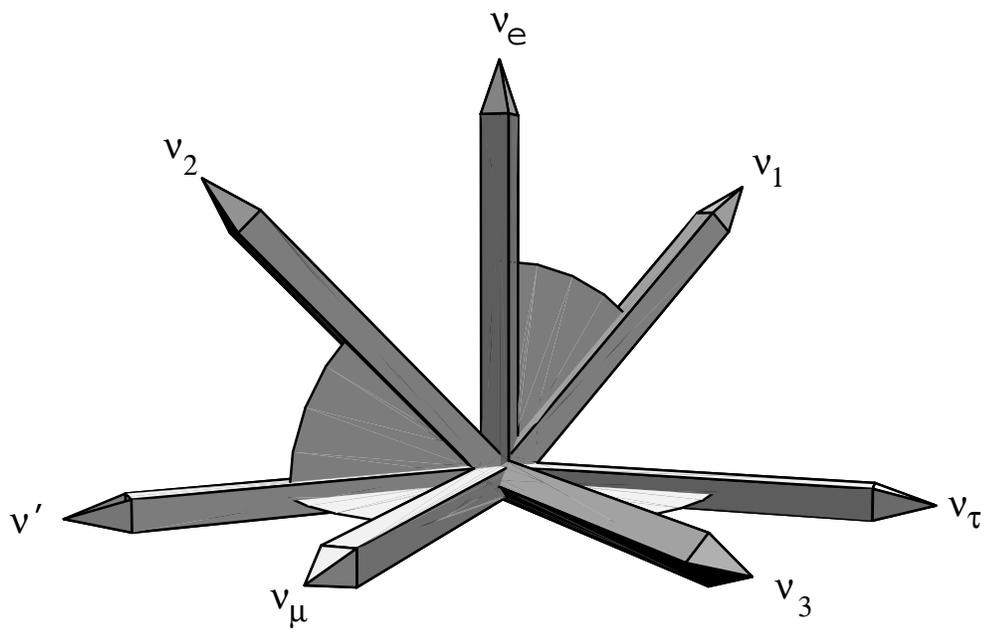
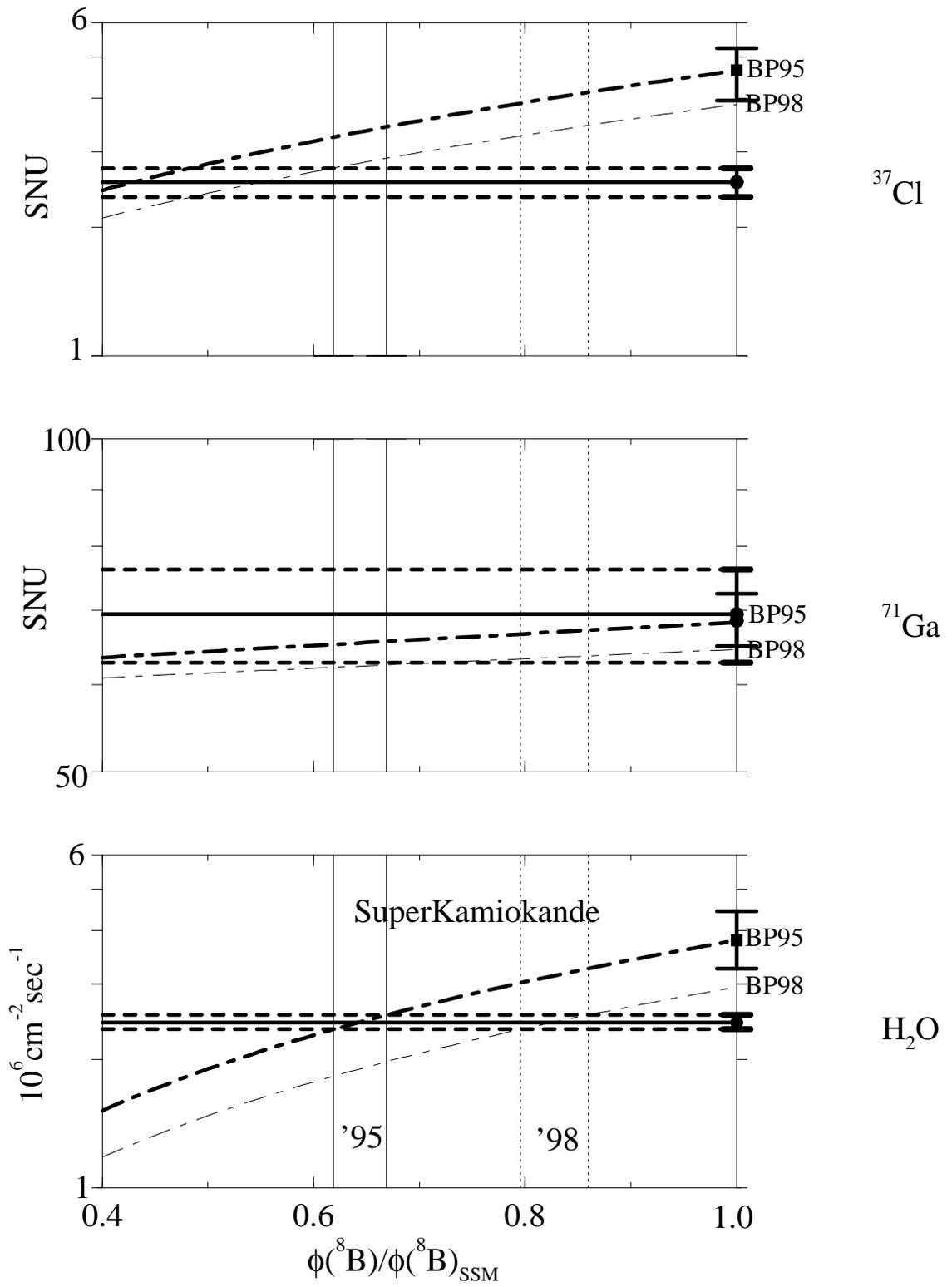
\end{document}